\documentclass[fleqn]{revtex4}

\usepackage{graphicx}
\usepackage{amsmath}
\usepackage{dcolumn}
\usepackage{rotating}
\usepackage{makecell}
\usepackage{longtable}
\usepackage{enumitem}
\usepackage[francais]{babel}
\usepackage[latin1]{inputenc}
\begin{document}

\title{Stark quenching of rovibrational states of H$_2^+$ due to motion in a magnetic field}

\author{Jean-Philippe Karr}
\affiliation{Laboratoire Kastler Brossel, Sorbonne Université, CNRS, ENS-PSL Research University, Collège de France\\
4 place Jussieu, F-75005 Paris, France}
\affiliation{Université d'Evry-Val d'Essonne, Université Paris-Saclay, Boulevard François Mitterrand, F-91000 Evry, France}

\begin{abstract}
The motional electric field experienced by an H$_2^+$ ion moving in a magnetic field induces an electric dipole, so that one-photon dipole transitions between rovibrational states become allowed. Field-induced spontaneous decay rates are calculated for a wide range of states. For an ion stored in a high-field ($B \sim 10$~T) Penning trap, it is shown that the lifetimes of excited rovibrational states can be shortened by typically 1-3 orders of magnitude by placing the ion in a large cyclotron orbit. This can greatly facilitate recently proposed [E. G.~Myers, Phys. Rev. A~\textbf{98}, 010101 (2018)] high-precision spectroscopic measurements on H$_2^+$ and its antimatter counterpart for tests of $CPT$ symmetry.
\end{abstract}

\maketitle

\section*{Introduction}

The hydrogen molecular ions, H$_2^+$ and its isotopes (HD$^+$, D$_2^+$) have long since been identified as a highly promising system for fundamental physics tests. High-precision spectroscopy of rovibrational transitions can be used to determine the proton-electron and deuteron-electron mass ratios~\cite{wing1976,roth2008} and probe their possible time
variation~\cite{schiller2005,schiller2014,karr2014}. Furthermore, measuring an appropriate set of transitions in H$_2^+$ and HD$^+$ would allow a determination of the proton and deuteron
charge radii and the Rydberg constant~\cite{karr2016} and thus may contribute to resolving the existing discrepancies on their values~\cite{antognini2013,arrington2015,pohl2016,beyer2017,fleurbaey2018}. Spectroscopy of H$_2^+$ compared with its antimatter counterpart $\bar{H}_2^-$ was recently proposed for improved tests of
$CPT$ symmetry~\cite{myers2018}.

Theoretical predictions of rovibrational transition frequencies have surpassed the 10$^{-11}$ accuracy level~\cite{korobov2017a} which already allows for an improved determination of mass
ratios~\cite{sturm2014,heisse2017}. In HD$^+$, experiments with ultracold trapped ions have reached a precision of 10$^{-9}$~\cite{bressel2012,biesheuvel2016} and more recently
$5\!\times\!10^{-10}$~\cite{alighanbari2018}. Studies on H$_2^+$~{\cite{jefferts1969,carrington1993,critchley2001,haase2015,beyer2016} have been so far performed by different techniques yielding lower accuracies. The main obstacle faced by ultrahigh-resolution spectroscopy of trapped H$_2^+$ (or $\bar{H}_2^-$) is quantum state preparation, due to the fact that, unlike in the heteronuclear HD$^+$, rovibrational transitions are not dipole-allowed. As a result, rovibrational states of vibration quantum number $v \geq 1$ have very long lifetimes typically of the order of 10 days~\cite{posen1983,pilon2012}. It would take several months for an ion produced in e.g. $v=10$ to decay down to $v=0$. Pure rotational transitions for lower values of rotational quantum number $L$ are even slower, the rotational levels of $v=0$ with $L<8$ having lifetimes greater than one year, for example. In addition, optical pumping methods for cooling the internal degrees of freedom~\cite{schneider2010} are not available.
One solution~\cite{karr2012} consists in creating the ions directly in the desired rovibrational state by state-selective multiphoton ionization (REMPI) of H$_2$~\cite{anderson1984,ohalloran1987}. However, no selective production scheme is available for antimatter $\bar{H}_2^-$ ions~\cite{myers2018}. Rovibrational cooling by cold He buffer gas has also been proposed~\cite{schiller2017} but is again not well suited for antimatter ions.

In this paper, we study another rovibrational relaxation process relevant for experiments performed in a Penning trap, first proposed in~\cite{myers2018}. The idea is to exploit the motional electric field experienced by an ion as it moves in a magnetic field. This electric field polarizes the ion~\cite{thompson2004} making one-photon ro-vibrational decay allowed. The Stark quenching process is much less efficient in H$_2^+$ as e.g. for the 2S state in hydrogen~\cite{lamb1950} because the mixing of 1s$\sigma_g$ rovibrational states induced by the electric field occurs with excited electronic states and is very weak. Nevertheless, our results show that rovibrational decay can typically be accelerated by at least one order of magnitude in a high-field ($B \sim 10$~T) Penning trap.

In the following, we present the calculation of decay rates for a wide range of rovibrational states. The paper is structured as follows: the theoretical expression of the decay rate is derived in Sec.~\ref{theory}. The numerical method is described in Sec.~\ref{method}. Results are presented in Sec.~\ref{results}, and their experimental implications briefly discussed.

\section{Theory} \label{theory}

\subsection{Motional electric field}

We consider an H$_2^+$ ion in a Penning trap with a uniform magnetic field $\mathbf{B_0}$ along the $z$ axis, placed in a cyclotron orbit of radius $r_c$. The ion's velocity is
\begin{equation}
\mathbf{v}(t) = \frac{qB_0r_c}{m} \left[ \sin(\omega_c t) \mathbf{e_x} + \cos (\omega_c t) \mathbf{e_y} \right]
\end{equation}
where $\omega_c = qB_0/m$ is the cyclotron frequency. The moving ion experiences a motional electric field
\begin{equation}
\mathbf{E_0}(t) = \mathbf{v}(t) \times \mathbf{B_0} = E_0 \left[ \cos(\omega_c t) \mathbf{e_x} - \sin (\omega_c t) \mathbf{e_y} \right]
\end{equation}
with $E_0 = qB_0^2r_c/m$.
This electric field can be expressed in terms of the standard polarizations
\begin{equation}
\boldsymbol{\epsilon}_{\pm 1} = \mp \frac{1}{\sqrt{2}} \left( \mathbf{e_x} \pm i \mathbf{e_y} \right).
\end{equation}
One gets:
\begin{equation}
\mathbf{E_0} (t) = \frac{E_0}{\sqrt{2}} \left( e^{-i\omega_c t} \boldsymbol{\epsilon}_{-1} + h.c. \right).
\end{equation}
Decay induced by the field $\mathbf{E_0} (t)$ can be understood as a two-photon process whereby a ''cyclotron'' photon of energy $\hbar \omega_c$ is emitted (or absorbed) and another is spontaneously emitted. Since the cyclotron frequency is very small with respect to rovibrational transition frequencies in H$_2^+$, one may approximate $\mathbf{E_0} (t)$ by a static field
\begin{equation}
\mathbf{E_0} = \frac{E_0}{\sqrt{2}} \left( \boldsymbol{\epsilon}_{-1} - \boldsymbol{\epsilon}_{1} \right).
\end{equation}

\subsection{Dipole matrix elements}

A ro-vibrational state $(v,L)$ of H$_2^+$ supported by the ground (1s$\sigma_g$) electronic curve will be mixed by the motional electric field with states of opposite parity supported by excited electronic curves. At the first order of perturbation theory, the perturbed wavefunction may be written as
\begin{equation}
\left| \psi_{v,L,M}^{(1)} \right\rangle= \left| \psi_{v,L,M} \right\rangle +  \frac{1}{E_{v,L}-H} \mathbf{d} \cdot \mathbf{E_0} \left| \psi_{v,L,M} \right\rangle
\end{equation}
As a result, one-photon transitions between ro-vibrational states become dipole-allowed. Keeping only leading-order terms, the transition dipole moment is
\begin{eqnarray}
\left\langle \psi_{v',L',M'}^{(1)} \right| \mathbf{d}\!\cdot\!\boldsymbol{\epsilon} \left| \psi_{v,L,M}^{(1)} \right\rangle &=& E_0 \left( \left\langle \psi_{v',L',M'} \right| \mathbf{d}\!\cdot\!\boldsymbol{\epsilon} \frac{1}{E_{v,L}-H} \mathbf{d} \!\cdot\! \boldsymbol{\epsilon}_0 \left| \psi_{v,L,M} \right\rangle \right. \nonumber \\ && \left. + \left\langle \psi_{v',L',M'} \right| \mathbf{d}\!\cdot\!\boldsymbol{\epsilon}_0 \frac{1}{E_{v',L'}-H} \mathbf{d}\!\cdot\!\boldsymbol{\epsilon} \left| \psi_{v,L,M} \right\rangle \right)
\end{eqnarray}
where $\boldsymbol{\epsilon}_0 = \left( \boldsymbol{\epsilon}_{-1} - \boldsymbol{\epsilon}_{1} \right)/\sqrt{2}$. This can be written in terms of the two-photon transition operator $Q_{\boldsymbol{\epsilon}_1\boldsymbol{\epsilon}_2}(\omega_1,\omega_2)$~\cite{grynberg1976,grynberg1977}:
\begin{equation}
\left\langle \psi_{v',L',M'}^{(1)} \right| \mathbf{d}\!\cdot\!\boldsymbol{\epsilon} \left| \psi_{v,L,M}^{(1)} \right\rangle = E_0 \left\langle \psi_{v',L',M'} \right| Q_{\boldsymbol{\epsilon}_0\boldsymbol{\epsilon}}(0,\omega) \left| \psi_{v,L,M} \right\rangle, \label{dipole}
\end{equation}
where $\hbar \omega = E_{v,L} - E_{v',L'}$, and
\begin{eqnarray}
Q_{\boldsymbol{\epsilon}_1\boldsymbol{\epsilon}_2}(\omega_1,\omega_2) &=& Q_{\boldsymbol{\epsilon}_2\boldsymbol{\epsilon}_1}(\omega_1) + Q_{\boldsymbol{\epsilon}_1\boldsymbol{\epsilon}_2}(\omega_2), \\
Q_{\boldsymbol{\epsilon}_2\boldsymbol{\epsilon}_1}(\omega_1) &=& \mathbf{d}\!\cdot\!\boldsymbol{\epsilon}_2 \frac{1}{E_{v,L}-\hbar\omega_1-H} \mathbf{d}\!\cdot\!\boldsymbol{\epsilon}_1
\end{eqnarray}
\subsection{Properties of the two-photon transition operator}

Since this operator couples twice the vector operator $\mathbf{d}$, it may be written as a sum of operators of rank 0, 1 and 2. Furthermore, one can decompose it into its symmetrical and antisymmetrical parts~\cite{grynberg1976}
\begin{eqnarray}
Q_{\boldsymbol{\epsilon}_1\boldsymbol{\epsilon}_2} (\omega_1,\omega_2) &=& Q_{\boldsymbol{\epsilon}_1\boldsymbol{\epsilon}_2}^S (\omega_1,\omega_2) + Q_{\boldsymbol{\epsilon}_1\boldsymbol{\epsilon}_2}^A (\omega_1,\omega_2), \\
Q_{\boldsymbol{\epsilon}_1\boldsymbol{\epsilon}_2}^S (\omega_1,\omega_2) &=& \frac{1}{2} \left( Q_{\boldsymbol{\epsilon}_1\boldsymbol{\epsilon}_2}(\omega_1,\omega_2) + Q_{\boldsymbol{\epsilon}_2\boldsymbol{\epsilon}_1}(\omega_1,\omega_2) \right) \\
Q_{\boldsymbol{\epsilon}_1\boldsymbol{\epsilon}_2}^A (\omega_1,\omega_2) &=& \frac{1}{2} \left( Q_{\boldsymbol{\epsilon}_1\boldsymbol{\epsilon}_2}(\omega_1,\omega_2) - Q_{\boldsymbol{\epsilon}_2\boldsymbol{\epsilon}_1}(\omega_1,\omega_2) \right).
\end{eqnarray}
Then $Q_{\boldsymbol{\epsilon}_1\boldsymbol{\epsilon}_2}^S (\omega_1,\omega_2)$ is decomposed into operators of ranks 0 and 2, while $Q_{\boldsymbol{\epsilon}_1\boldsymbol{\epsilon}_2}^A (\omega_1,\omega_2)$ is of rank 1.

Selection rules for Stark-induced decay are thus identical to those of two-photon transitions~\cite{karr2008}, i.e. $\Delta L = 0,\pm2$ for rovibrational transitions within the 1s$\sigma_g$ electronic state. $L = 0 \to L' = 0$ transitions only involve the scalar component of the two-photon operator, while $L \to L \pm 2$ transitions only involve the rank 2 component. The $L \to L$
transitions with $L \geq 1$ are the most complicated cases, involving all three terms (rank 0, 1 and 2).

We define the scalar ($Q^{(0)}$) and tensor ($Q^{(2)}$) operators appearing in the decomposition of $Q_{\boldsymbol{\epsilon}_1\boldsymbol{\epsilon}_2}^S$ by the relationship
\begin{equation}
Q_{00}^S = Q^{(0)}_0 + Q^{(2)}_0
\end{equation}
The required $Q_{q_1 q_2}^S$ components can then be expressed in terms of irreducible components $Q^{(0)}_q$, $Q^{(2)}_q$. One obtains~\cite{karr2014}
\begin{eqnarray}
Q_{\pm 1\pm 1}^S &=& \sqrt{\frac{3}{2}} Q_{\pm 2}^{(2)} \\
Q_{\pm 1 0}^S &=& \frac{\sqrt{3}}{2} Q_{\pm 1}^{(2)} \\
Q_{\pm 1\mp 1}^S &=& -Q^{(0)}_0 + \frac{1}{2} Q_0^{(2)}
\end{eqnarray}
Similarly, for the antisymmetrical part $Q_{\boldsymbol{\epsilon}_1\boldsymbol{\epsilon}_2}^A$ we define the vector operator $Q^{(1)}$ by the relationship
\begin{equation}
Q_{-10}^A = Q_{-1}^{(1)}
\end{equation}
and the $Q_{q_1 q_2}^A$ components can be expressed in terms of irreducible components $Q^{(1)}_q$~\cite{grynberg1976}
\begin{eqnarray}
Q^A_{\pm 1 0} &=& \mp Q^{(1)}_{\pm 1} \\
Q^A_{\pm 1 \mp 1} &=& \mp Q^{(1)}_0
\end{eqnarray}
Finally, we set
\begin{eqnarray}
Q_s &=& \frac{ \left\langle v'L' || Q^{(0)} || vL \right\rangle }{\sqrt{2L'+1}} \\
Q_v &=& \frac{ \left\langle v'L' || Q^{(1)} || vL \right\rangle }{\sqrt{2L'+1}} \\
Q_t &=& \frac{ \left\langle v'L' || Q^{(2)} || vL \right\rangle }{\sqrt{2L'+1}}.
\end{eqnarray}

\subsection{Decay rates}

The decay rate (or Einstein coefficient $A$) of a given state $(v,L,M)$ associated with a transition towards a state $(v',L')$ (in s$^{-1}$) is given by
\begin{equation}
A_{v,L,M,v',L'} = \frac{2 \omega^3}{3 \epsilon_0 hc^3} \; \sum_{M',\boldsymbol{\epsilon}} \left| \left\langle \psi_{v'L'M'} \right| \mathbf{d\!\cdot\!\boldsymbol{\epsilon}} \left| \psi_{v,L,M} \right\rangle \right|^2 \label{einstein}
\end{equation}
In the usual case of spontaneous emission in an isotropic environment, the decay rate does not depend on $M$, but here anisotropy arises from the fact that the electric field $\mathbf{E_0}$ is polarized perpendicularly to the $z$ axis.

\noindent In the simplest case $L = L' = 0$ we get
\begin{equation}
A_{v,L=0,M=0,v',L'=0} = \frac{2 \omega^3}{3 \epsilon_0 hc^3} E_0^2 Q_s^2. \label{rate-l0}
\end{equation}
For $L' = L \pm 2$:
\begin{eqnarray}
A_{v,L,M,v',L'} &=& \frac{2 \omega^3}{3 \epsilon_0 hc^3} \frac{E_0^2}{2} \sum_{M',q} \left( \left| \left\langle v'L'M' \right| Q_{-1q}^S(0,\omega) \left| v L M \right\rangle \right|^2  + \left| \left\langle v'L'M' \right| Q_{1q}^S(0,\omega) \left| v L M \right\rangle \right|^2 \right) \nonumber \\
&=& \frac{\omega^3}{3 \epsilon_0 hc^3} E_0^2  \left\{ \frac{3}{2} \left| \left\langle v'L'M\!+\!2 \right| Q_2^{(2)} \left| vLM \right\rangle \right|^2 + \frac{3}{2} \left| \left\langle v'L'M\!-\!2 \right| Q_{-2}^{(2)} \left| vLM \right\rangle \right|^2 \right. \nonumber \\
&& \left. + \frac{3}{4} \left| \left\langle v'L'M\!+\!1 \right| Q_1^{(2)} \left| vLM \right\rangle \right|^2 + \frac{3}{4} \left| \left\langle v'L'M\!-\!1 \right| Q_{-1}^{(2)} \left| vLM \right\rangle \right|^2 + \frac{1}{2}\left| \left\langle v'L'M \right| Q_0^{(2)} \left| vLM \right\rangle \right|^2 \right\} \nonumber \\
&=& \frac{\omega^3}{3 \epsilon_0 hc^3} E_0^2  \left\{ \frac{3}{2} \left| \left\langle L M 2 2 | L' M\!+\!2 \right\rangle \right|^2 + \frac{3}{2} \left| \left\langle L M 2 -2 | L' M\!-\!2 \right\rangle \right|^2 \right.  \label{rate-dl2} \\
&& \left. + \frac{3}{4} \left| \left\langle L M 2 1 | L' M\!+\!1 \right\rangle \right|^2 + \frac{3}{4} \left| \left\langle L M 2 -1 | L' M\!-\!1 \right\rangle \right|^2 + \frac{1}{2} \left| \left\langle L M 2 0 | L' M \right\rangle \right|^2 \right\} Q_t^2 \nonumber
\end{eqnarray}
One can show that the averaged decay rate $\bar{A}_{v,L,v',L'} = \left(\sum_{M=-L}^L A_{v,L,M,v',L'} \right)/(2L+1)$ is
\begin{equation}
\bar{A}_{v,L,v',L'} = \frac{\omega^3}{3 \epsilon_0 hc^3} E_0^2 Q_t^2 \frac{2 L' + 1}{2 L + 1}. \label{avg-rate-dl2}
\end{equation}

Finally, for $L' = L \geq 1$ the spontaneous emission rate is
\begin{eqnarray}
A_{v,L,M,v',L} &=& \frac{2 \omega^3}{3 \epsilon_0 hc^3} E_0^2  \sum_{M'} \left| \left\langle v' L M' | Q_{\boldsymbol{\epsilon}_0\boldsymbol{\epsilon}}^S (0,\omega) | v L M \right\rangle + \left\langle v' L M' | Q_{\boldsymbol{\epsilon}_0\boldsymbol{\epsilon}}^A (0,\omega) | v L M \right\rangle \right|^2 \nonumber \\
&=& \frac{2 \omega^3}{3 \epsilon_0 hc^3} \frac{E_0^2}{2} \sum_{M',q} \left| \left\langle v'L M' \right| Q_{-1q}^S \left| v L M \right\rangle + \left\langle v'L M' \right| Q_{1q}^S \left| v L M \right\rangle \right. \nonumber \\
&& \left. + \left\langle v'L M' \right| Q_{-1q}^A \left| v L M \right\rangle + \left\langle v'L M' \right| Q_{1q}^A \left| v L M \right\rangle \right|^2 \nonumber \\
&=& \frac{\omega^3}{3 \epsilon_0 hc^3} E_0^2 \left\{ \frac{3}{2} \left(\left| \left\langle L M 2 2 | L M\!+\!2 \right\rangle \right|^2 + \left| \left\langle L M 2 -2 | L M\!-\!2 \right\rangle  \right|^2 \right) Q_t^2 \right. \label{rate-dl0}\\
&& + \left| \frac{\sqrt{3}}{2} \left\langle L M 2 1 | L M\!+\!1 \right\rangle Q_t - \left\langle L M 1 1 | L M\!+\!1 \right\rangle Q_v \right|^2 \nonumber \\
&& + \left| \frac{\sqrt{3}}{2} \left\langle L M 2 -1 | L M\!-\!1 \right\rangle Q_t + \left\langle L M 1 -1 | L M\!-\!1 \right\rangle Q_v \right|^2 \nonumber \\
&& + \left| -Q_s + \frac{1}{2} \left\langle L M 2 0 | L M \right\rangle Q_t + \left\langle L M 1 0 | L M \right\rangle Q_v \right|^2 \nonumber \\
&& \left. + \left| -Q_s + \frac{1}{2} \left\langle L M 2 0 | L M \right\rangle Q_t - \left\langle L M 1 0 | L M \right\rangle Q_v \right|^2 \right\} \nonumber
\end{eqnarray}
The averaged decay rate is
\begin{equation} \label{avg-rate-dl0}
\bar{A}_{v,L,v',L} = \frac{\omega^3}{3 \epsilon_0 hc^3} E_0^2 \left( 2 Q_s^2 + Q_t^2 + \frac{4}{3} Q_v^2 \right).
\end{equation}

\section{Numerical method} \label{method}

We now present the calculation of two-photon reduced matrix elements. Our numerical method has been previously described in Refs.~\cite{karr2014,karr2008,korobov2017b}, and we briefly recall the main points here. The three-body Schr\"odinger equation is solved using the following variational expansion of the wave function:
\begin{equation}
\begin{array}{@{}l}
\displaystyle \Psi_{LM}(\mathbf{r}_1,\mathbf{r}_2) =
       \sum_{l_1+l_2=L}
         \mathcal{Y}^{l_1l_2}_{LM}(\mathbf{r}_1,\mathbf{r}_2)
         R^{L}_{l_1l_2}(r_1,r_2,r_{12}),
\\[4mm]\displaystyle
R_{l_1l_2}^{L}(r_1,r_2,r_{12}) = \sum_{n=1}^{N_{l_1}} \Big\{C_n\,\mbox{Re} \bigl[e^{-\alpha_n r_{12}-\beta_n r_1-\gamma_n r_2}\bigr]
+D_n\,\mbox{Im} \bigl[e^{-\alpha_n r_{12}-\beta_n r_1-\gamma_n r_2}\bigr] \Big\}.
\end{array}
\end{equation}
$r_1,r_2,r_{12}$ are the interparticle distances, and $\mathcal{Y}^{l_1l_2}_{LM}(\mathbf{r}_1,\mathbf{r}_2) = r_1^{l_1} r_2^{l_2} Y^{l_1l_2}_{LM}(\hat{\mathbf{r}}_1,\hat{\mathbf{r}}_2)$ where $Y^{l_1l_2}_{LM}$ is a bipolar spherical harmonic. The complex exponents $\alpha_n$, $\beta_n$, $\gamma_n$ are generated pseudorandomly in several intervals. All parameters, i.e. interval bounds and the number of basis functions $N_{i,l_1}$ in each interval $i$ and angular momentum subset $\{l_1,l_2\}$  (keeping the total basis length $N = 2\sum_{i,l_1} N_{i,l_1}$ constant), have been optimized for a few tens of states. This yielded accuracies of $10^{-12}$ a.u. or better on the energy levels of all 201 rovibrational levels considered here using basis lengths $N = 2000-3500$.

For the calculation of two-photon matrix elements, the following three terms, which correspond to the possible values $L'\!-\!1$, $L'\!+\!1$, $L'$ for the angular momentum of the intermediate state, are evaluated numerically:
\begin{equation}
a_{0,\pm} = \frac{1}{\sqrt{(2L+1)(2L'+1)}} \left[ \langle v'L' \| \mathbf{d} (E_{vL}-H_{L''})^{-1} \mathbf{d} \| vL \rangle + \langle v'L' \| \mathbf{d} (E_{v'L'}-H_{L''})^{-1} \mathbf{d} \| vL \rangle \right]
\end{equation}
where $H_{L''}$ denotes the restriction of the Schr\"odinger Hamiltonian $H$ to a subspace of angular momentum $L''$, and $a_0$, $a_+$, $a_-$ respectively correspond to $L''=L',L'+1,L'-1$. Basis lengths $N_{L''} = 2000-3000$ are typically used for $H_{L''}$ which is more than sufficient to obtain these quantities with five significant digits. The reduced matrix elements of $Q^{(k)}$ are related to $a_-$, $a_+$, $a_0$ in the following way:
\begin{eqnarray}
\frac{\left\langle \!v'\!L\|Q^{(0)}\|v\!L\!\right\rangle}{\sqrt{2L+1}} &=& \frac{1}{3}\>\bigl(a_-+a_0+a_+\bigr)\>\\
\frac{\left\langle \!v'\!L\|Q^{(1)}\|v\!L\!\right\rangle}{\sqrt{2L+1}} &=& \sqrt{L(L\!+\!1)}\>\left[\frac{a_-}{L}+\frac{a_0}{L(L\!+\!1)}-\frac{a_+}{L\!+\!1}\right]\>\\
\frac{\left\langle \!v'\!L\!+\!2\|Q^{(2)}\|v\!L\!\right\rangle}{\sqrt{2L+5}} &=& -\sqrt{\frac{2(2L\!+\!1)}{3(2L\!+\!3)}}\>a_-\>\\
\frac{\left\langle \!v'\!L\|Q^{(2)}\|v\!L\!\right\rangle}{\sqrt{2L+1}} &=& -\frac{1}{3}\sqrt{(2L\!+\!3)(2L\!-\!1)L(L\!+\!1)}\! \left[\frac{a_-}{L(2L\!-\!1)} - \frac{a_0}{L(L\!+\!1)}+ \frac{a_+}{(2L\!+\!3)(L\!+\!1)}\right]\\
\frac{\left\langle \!v'\!L\!-\!2\|Q^{(2)}\|v\!L\!\right\rangle}{\sqrt{2L-3}}&=&-\sqrt{\frac{2(2L\!+\!1)}{3(2L\!-\!1)}}\;a_+\>
\end{eqnarray}
\section{Results and discussion} \label{results}
The range of ro-vibrational states was chosen in order to provide all the results relevant to any of the following experimental situations:
\begin{enumerate}[label=(\roman*)]
\item H$_2^+$ ions produced by electron-impact ionization of H$_2$. This process has been shown to create ions predominantly in $v=0-12$, $L=0-4$ with only $\sim 1$ percent probability of populating higher vibrational states~\cite{vonbusch1972}.
\item $\bar{H}_2^-$ antimatter ions produced through the reaction $\bar{H}^+ + \bar{p} \to \bar{H}_2^- + e^+$, as proposed in~\cite{myers2018}, using $\bar{H}^+$ ions to be produced in the GBAR experiment~\cite{perez2015}. The reaction being exothermic by 1.896~eV, the $\bar{H}_2^-$ ion will be produced with $v=0-8$ and $L=0-27$, with low-$v$, high $L$ states being favored.
\end{enumerate}
The two-photon matrix elements $Q_s$, $Q_v$ and $Q_t$ were calculated for all the required states~\cite{suppl}. The corresponding $M$-averaged decay rates $A/(B_0^4 r_c^2)$ in s$^{-1}$T$^{-4}$mm$^{-2}$ are given in the following Tables for $L \to L' = L-2$ ($\Delta v = v - v' = 0,1,2$, Tables~\ref{A-dlm2-dv0}-\ref{A-dlm2-dv2}), $L \to L' = L$ ($\Delta v = 1,2$, Tables~\ref{A-dl0-dv1}-\ref{A-dl0-dv2}), and $L \to L' = L+2$ transitions ($\Delta v = 1,2$, Tables~\ref{A-dlp2-dv1}-\ref{A-dlp2-dv2}). The total decay rates $\bar{A}_{v,L} = \sum_{v',L'} \bar{A}_{v,L,v',L'}$ are given in Table~\ref{A-total}.

The decay rates decrease sharply with increasing $\Delta v$, as previously observed in the case of two-photon transitions~\cite{hilico2001}. Decay towards $v' =  v - 3$ states was also included in the total rates of Table~\ref{A-total}. However, since the fractional contribution of $\Delta v = 3$ transitions to the overall decay rates amounts to less than 10$^{-3}$ in all cases, these decay rates are not reported here.

The dependence of the decay rates on the magnetic quantum number $M$ is illustrated in Fig.~\ref{rate-M}. One can see that the decay rate increases with $|M|$.

\begin{figure}[!ht]
\begin{center}
\includegraphics[width=8cm]{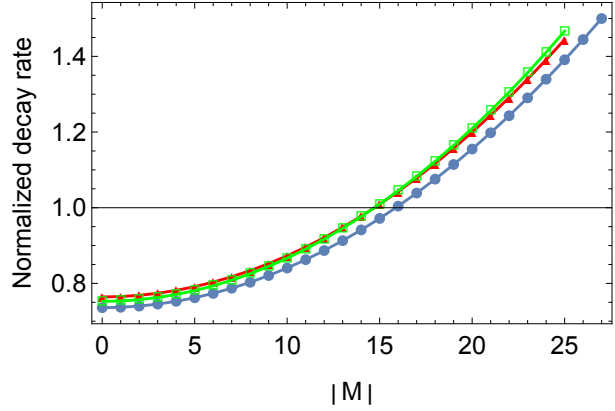}
\caption{Normalized decay rate $A_{v,L,M,v',L'}/\bar{A}_{v,L,v',L'}$ for $L=27,L'=25$ (blue circles), $L=25,L'=27$ (red triangles), and $v=1,L=25,v'=0,L'=25$ (green squares). Note that the
$M$ dependence is independent of $v$ and $v'$ for $\Delta L = \pm 2$ (see Eq.~(\ref{rate-dl2})), but not for $\Delta L = 0$ (Eq.~(\ref{rate-dl0})).}
\label{rate-M}
\end{center}
\end{figure}

These results can be used to precisely model the time evolution of $H_2^+$ (or $\bar{H}_2^-$) ro-vibrational populations from initial conditions corresponding to different production schemes. Such a study is outside the scope of the present paper, but it is already possible to draw some qualitative conclusions. Fig.~\ref{rate-ex} shows both zero-field and Stark-quenched lifetimes of selected ro-vibrational states, for a magnetic field value $B_0 = 8.5$~T and a cyclotron radius $r_c = 2$~mm which correspond to the parameters of the Penning trap operated in Tallahassee~\cite{smith2018}.

\begin{figure}[!ht]
\begin{center}
\includegraphics[width=8cm]{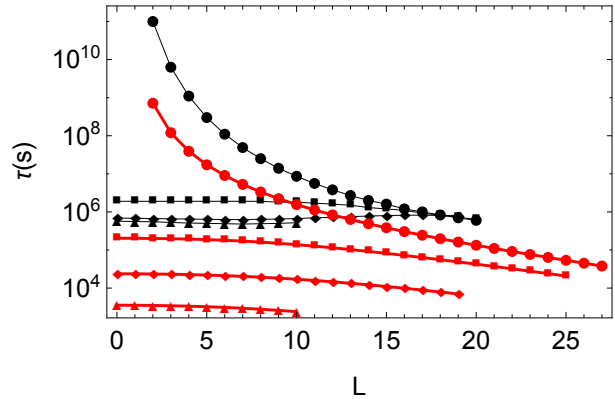}
\caption{Stark lifetimes of H$_2^+$ rovibrational states for an ion in a cyclotron orbit of radius $r_c = 2$~mm in a magnetic field $B_0 = 8.5$~T are shown by thick red lines. For comparison, the zero-field lifetimes (taken from~\cite{posen1983}) are shown by thin black lines. Circles, squares, diamonds and triangles respectively stand for $v=0,1,4,8$.}
\label{rate-ex}
\end{center}
\end{figure}

One may observe that Stark quenching is most efficient for excited vibrational states, leading to lifetimes of e.g. less than one hour for $v=8$, a few hours for $v=4$, and a few days for $v=1$. This represents a reduction by 1-2 orders of magnitude with respect to the zero-field lifetime. Quenching of the vibrational motion down to $v=0$ may thus be expected to occur within~$\sim 1$ week in present-day Penning traps. Quenching of the rotational motion is significantly slower, with lifetimes ranging from $\sim 10$~hours for $(v=0,L=27)$ up to $\sim 23$~years for $(v=0,L=2)$. Thus for manipulation of antimatter molecular ions which may be formed in high-$L$ states~\cite{myers2018} it would be useful to transfer the ions to a larger trap where the ions could be placed in a cyclotron orbit of very large radius. For example, for $B_0 = 10$~T and $r_c = 20$~mm, the lifetimes would range from $\sim 3$~minutes for $(v=0,L=27)$ to $\sim 7$~days for $(v=0,L=3)$. Rotational quenching down to $L=1$ or $2$ could thus occur in about 1 week.

In conclusion, our results indicate that vibrational quenching of H$_2^+$ or its antimatter counterpart may be achieved within~$\sim 1$ week in existing Penning trap apparatus, and that rotational quenching is achievable over a similar time scale in a specially designed apparatus allowing for larger cyclotron radii. This opens new possibilities for ultra-high resolution spectroscopy of the simplest (anti)-molecule. Rovibrational Stark quenching of H$_2^+$ (and other homonuclear diatomic molecular ions) could also be significant if they are stored in heavy-ion storage rings in which the bending sections have comparably large values of $B_0^4 r_c^2$.

Finally, it is worth noting that, although we have focused on the case of an H$_2^+$ ion orbiting in an external magnetic field, the results presented here may readily be applied to calculate its decay rates in an external static electric field $\mathbf{E_0}$. Since the decay rate of a particular $M$ state depends on the electric field polarization, Eqs.~(\ref{rate-dl2}) and (\ref{rate-dl0}) have to be adapted to the specific case under study. However, the averaged decay rates are independent of the field polarization, so that Eqs.~(\ref{rate-l0}), (\ref{avg-rate-dl2}), and (\ref{avg-rate-dl0}) may be directly applied. The results of Tables~\ref{A-dlm2-dv0}-\ref{A-total} can also be directly exploited using the relationship $E_0 = qB_0^2r_c/m$.

\begin{acknowledgements}
I thank E.G. Myers for bringing this problem to my attention and for stimulating discussions. I also thank him, as well as L. Hilico, for useful suggestions on the manuscript. I acknowledge support as a fellow of the Institut Universitaire de France.
\end{acknowledgements}

\begin{sidewaystable}
\begin{center}
\caption{Averaged decay rates $\bar{A}_{v,L,v',L'}/(B_0^4 r_c^2)$ in s$^{-1}$T$^{-4}$mm$^{-2}$ with $v'=v$, $L'=L-2$.
\label{A-dlm2-dv0}}
\begin{tabular}{c@{\hspace{2mm}}l@{\hspace{2mm}}l@{\hspace{2mm}}l@{\hspace{2mm}}l@{\hspace{2mm}}l@{\hspace{2mm}}l@{\hspace{2mm}}l@{\hspace{2mm}}l@{\hspace{2mm}}l@{\hspace{2mm}}l@{\hspace{2mm}}l@{\hspace{2mm}}l@{\hspace{1mm}}l}
\hline\hline
$L/v$ & \makecell[c]{0} & \makecell[c]{1} & \makecell[c]{2} & \makecell[c]{3} & \makecell[c]{4} & \makecell[c]{5} & \makecell[c]{6} & \makecell[c]{7} & \makecell[c]{8} & \makecell[c]{9} & \makecell[c]{10} & \makecell[c]{11} & \makecell[c]{12} \\
\hline
 2 & 6.693E-14 & 1.154E-13 & 1.911E-13 & 3.085E-13 & 4.896E-13 & 7.703E-13 & 1.209E-12 & 1.904E-12 & 3.029E-12 & 4.900E-12 & 8.124E-12 & 1.395E-11 & 2.517E-11 \\
 3 & 3.996E-13 & 6.882E-13 & 1.139E-12 & 1.838E-12 & 2.918E-12 & 4.591E-12 & 7.207E-12 & 1.136E-11 & 1.808E-11 & 2.928E-11 & 4.861E-11 & 8.366E-11 &           \\
 4 & 1.224E-12 & 2.105E-12 & 3.482E-12 & 5.615E-12 & 8.912E-12 & 1.402E-11 & 2.203E-11 & 3.474E-11 & 5.537E-11 & 8.981E-11 & 1.495E-10 & 2.580E-10 &           \\
 5 & 2.774E-12 & 4.765E-12 & 7.875E-12 & 1.269E-11 & 2.014E-11 & 3.170E-11 & 4.983E-11 & 7.868E-11 & 1.256E-10 & 2.042E-10 & 3.409E-10 &           &           \\
 6 & 5.299E-12 & 9.084E-12 & 1.500E-11 & 2.416E-11 & 3.833E-11 & 6.035E-11 & 9.494E-11 & 1.501E-10 & 2.402E-10 & 3.916E-10 & 6.564E-10 &           &           \\
 7 & 9.062E-12 & 1.550E-11 & 2.556E-11 & 4.114E-11 & 6.527E-11 & 1.028E-10 & 1.620E-10 & 2.566E-10 & 4.116E-10 & 6.735E-10 &           &           &           \\
 8 & 1.435E-11 & 2.448E-11 & 4.031E-11 & 6.485E-11 & 1.029E-10 & 1.623E-10 & 2.559E-10 & 4.064E-10 & 6.539E-10 & 1.075E-09 &           &           &           \\
\hline\hline
\end{tabular}
\end{center}
\end{sidewaystable}

\begin{sidewaystable}
\begin{center}
\caption{Same as Table~\ref{A-dlm2-dv0}, with $v' = v-1$.
\label{A-dlm2-dv1}}
\begin{tabular}{c@{\hspace{2mm}}l@{\hspace{2mm}}l@{\hspace{2mm}}l@{\hspace{2mm}}l@{\hspace{2mm}}l@{\hspace{2mm}}l@{\hspace{2mm}}l@{\hspace{2mm}}l@{\hspace{2mm}}l@{\hspace{2mm}}l@{\hspace{2mm}}l@{\hspace{1mm}}l}
\hline\hline
$L/v$ & \makecell[c]{1} & \makecell[c]{2} & \makecell[c]{3} & \makecell[c]{4} & \makecell[c]{5} & \makecell[c]{6} & \makecell[c]{7} & \makecell[c]{8} & \makecell[c]{9} & \makecell[c]{10} & \makecell[c]{11} & \makecell[c]{12} \\
\hline
 2 & 2.819E-11 & 7.437E-11 & 1.476E-10 & 2.618E-10 & 4.384E-10 & 7.110E-10 & 1.134E-09 & 1.798E-09 & 2.857E-09 & 4.584E-09 & 7.489E-09 & 1.259E-08 \\
 3 & 3.959E-11 & 1.042E-10 & 2.066E-10 & 3.659E-10 & 6.122E-10 & 9.926E-10 & 1.583E-09 & 2.511E-09 & 3.990E-09 & 6.408E-09 & 1.048E-08 &           \\
 4 & 4.789E-11 & 1.258E-10 & 2.489E-10 & 4.404E-10 & 7.362E-10 & 1.193E-09 & 1.903E-09 & 3.019E-09 & 4.802E-09 & 7.721E-09 & 1.266E-08 &           \\
 5 & 5.512E-11 & 1.445E-10 & 2.855E-10 & 5.044E-10 & 8.426E-10 & 1.365E-09 & 2.177E-09 & 3.455E-09 & 5.502E-09 & 8.860E-09 &           &           \\
 6 & 6.198E-11 & 1.621E-10 & 3.197E-10 & 5.643E-10 & 9.418E-10 & 1.525E-09 & 2.434E-09 & 3.865E-09 & 6.162E-09 & 9.943E-09 &           &           \\
 7 & 6.876E-11 & 1.794E-10 & 3.533E-10 & 6.228E-10 & 1.039E-09 & 1.682E-09 & 2.685E-09 & 4.269E-09 & 6.817E-09 &           &           &           \\
 8 & 7.564E-11 & 1.970E-10 & 3.871E-10 & 6.818E-10 & 1.137E-09 & 1.841E-09 & 2.940E-09 & 4.680E-09 & 7.488E-09 &           &           &           \\
\hline\hline
\end{tabular}
\end{center}
\end{sidewaystable}

\begin{sidewaystable}
\begin{center}
\caption{Same as Table~\ref{A-dlm2-dv0}, with $v' = v-2$.
\label{A-dlm2-dv2}}
\begin{tabular}{c@{\hspace{2mm}}l@{\hspace{2mm}}l@{\hspace{2mm}}l@{\hspace{2mm}}l@{\hspace{2mm}}l@{\hspace{2mm}}l@{\hspace{2mm}}l@{\hspace{2mm}}l@{\hspace{2mm}}l@{\hspace{2mm}}l@{\hspace{1mm}}l}
\hline\hline
$L/v$ & \makecell[c]{2} & \makecell[c]{3} & \makecell[c]{4} & \makecell[c]{5} & \makecell[c]{6} & \makecell[c]{7} & \makecell[c]{8} & \makecell[c]{9} & \makecell[c]{10} & \makecell[c]{11} & \makecell[c]{12} \\
\hline
 2 & 1.874E-12 & 6.880E-12 & 1.704E-11 & 3.562E-11 & 6.787E-11 & 1.223E-10 & 2.129E-10 & 3.628E-10 & 6.114E-10 & 1.028E-09 & 1.739E-09 \\
 3 & 2.072E-12 & 7.608E-12 & 1.886E-11 & 3.945E-11 & 7.522E-11 & 1.356E-10 & 2.362E-10 & 4.024E-10 & 6.777E-10 & 1.138E-09 &           \\
 4 & 1.958E-12 & 7.197E-12 & 1.786E-11 & 3.741E-11 & 7.143E-11 & 1.290E-10 & 2.247E-10 & 3.829E-10 & 6.445E-10 & 1.080E-09 &           \\
 5 & 1.745E-12 & 6.426E-12 & 1.598E-11 & 3.355E-11 & 6.418E-11 & 1.160E-10 & 2.024E-10 & 3.451E-10 & 5.804E-10 &           &           \\
 6 & 1.503E-12 & 5.550E-12 & 1.385E-11 & 2.915E-11 & 5.591E-11 & 1.013E-10 & 1.770E-10 & 3.018E-10 & 5.073E-10 &           &           \\
 7 & 1.260E-12 & 4.674E-12 & 1.171E-11 & 2.475E-11 & 4.764E-11 & 8.657E-11 & 1.515E-10 & 2.584E-10 &           &           &           \\
 8 & 1.032E-12 & 3.849E-12 & 9.697E-12 & 2.060E-11 & 3.983E-11 & 7.262E-11 & 1.273E-10 & 2.172E-10 &           &           &           \\
\hline\hline
\end{tabular}
\end{center}
\end{sidewaystable}

\begin{sidewaystable}
\begin{center}
\caption{Averaged decay rates $\bar{A}_{v,L,v',L'}/(B_0^4 r_c^2)$ in s$^{-1}$T$^{-4}$mm$^{-2}$ with $v'=v-1$, $L'=L$.
\label{A-dl0-dv1}}
\begin{tabular}{c@{\hspace{2mm}}l@{\hspace{2mm}}l@{\hspace{2mm}}l@{\hspace{2mm}}l@{\hspace{2mm}}l@{\hspace{2mm}}l@{\hspace{2mm}}l@{\hspace{2mm}}l@{\hspace{2mm}}l@{\hspace{2mm}}l@{\hspace{2mm}}l@{\hspace{1mm}}l}
\hline\hline
$L/v$ & \makecell[c]{1} & \makecell[c]{2} & \makecell[c]{3} & \makecell[c]{4} & \makecell[c]{5} & \makecell[c]{6} & \makecell[c]{7} & \makecell[c]{8} & \makecell[c]{9} & \makecell[c]{10} & \makecell[c]{11} & \makecell[c]{12} \\
\hline
0  & 1.298E-10 & 3.033E-10 & 5.405E-10 & 8.716E-10 & 1.343E-09 & 2.027E-09 & 3.040E-09 & 4.576E-09 & 6.963E-09 & 1.079E-08 & 1.713E-08 & 2.816E-08 \\
1  & 1.794E-10 & 4.344E-10 & 8.014E-10 & 1.335E-09 & 2.120E-09 & 3.289E-09 & 5.055E-09 & 7.771E-09 & 1.204E-08 & 1.893E-08 & 3.044E-08 & 5.052E-08 \\
2  & 1.665E-10 & 4.003E-10 & 7.335E-10 & 1.215E-09 & 1.918E-09 & 2.962E-09 & 4.536E-09 & 6.952E-09 & 1.075E-08 & 1.688E-08 & 2.713E-08 & 4.505E-08 \\
3  & 1.659E-10 & 3.986E-10 & 7.301E-10 & 1.209E-09 & 1.909E-09 & 2.948E-09 & 4.516E-09 & 6.928E-09 & 1.072E-08 & 1.687E-08 & 2.717E-08 &           \\
4  & 1.673E-10 & 4.024E-10 & 7.376E-10 & 1.222E-09 & 1.932E-09 & 2.988E-09 & 4.584E-09 & 7.045E-09 & 1.093E-08 & 1.723E-08 & 2.785E-08 &           \\
5  & 1.698E-10 & 4.089E-10 & 7.506E-10 & 1.246E-09 & 1.973E-09 & 3.056E-09 & 4.699E-09 & 7.239E-09 & 1.126E-08 & 1.782E-08 &           &           \\
6  & 1.731E-10 & 4.175E-10 & 7.678E-10 & 1.277E-09 & 2.026E-09 & 3.147E-09 & 4.852E-09 & 7.497E-09 & 1.170E-08 & 1.860E-08 &           &           \\
\hline\hline
\end{tabular}
\end{center}
\end{sidewaystable}

\begin{sidewaystable}
\begin{center}
\caption{Same as Table~\ref{A-dl0-dv1}, with $v' = v-2$.
\label{A-dl0-dv2}}
\begin{tabular}{c@{\hspace{2mm}}l@{\hspace{2mm}}l@{\hspace{2mm}}l@{\hspace{2mm}}l@{\hspace{2mm}}l@{\hspace{2mm}}l@{\hspace{2mm}}l@{\hspace{2mm}}l@{\hspace{2mm}}l@{\hspace{2mm}}l@{\hspace{1mm}}l}
\hline\hline
$L/v$ & \makecell[c]{2} & \makecell[c]{3} & \makecell[c]{4} & \makecell[c]{5} & \makecell[c]{6} & \makecell[c]{7} & \makecell[c]{8} & \makecell[c]{9} & \makecell[c]{10} & \makecell[c]{11} & \makecell[c]{12} \\
\hline
0  & 1.717E-12 & 7.717E-12 & 2.265E-11 & 5.468E-11 & 1.178E-10 & 2.360E-10 & 4.500E-10 & 8.297E-10 & 1.497E-09 & 2.669E-09 & 4.748E-09 \\
1  & 6.366E-12 & 2.480E-11 & 6.501E-11 & 1.433E-10 & 2.867E-10 & 5.404E-10 & 9.804E-10 & 1.735E-09 & 3.026E-09 & 5.249E-09 & 9.135E-09 \\
2  & 5.116E-12 & 2.024E-11 & 5.376E-11 & 1.199E-10 & 2.424E-10 & 4.613E-10 & 8.435E-10 & 1.503E-09 & 2.638E-09 & 4.600E-09 & 8.045E-09 \\
3  & 5.001E-12 & 1.985E-11 & 5.291E-11 & 1.183E-10 & 2.399E-10 & 4.574E-10 & 8.381E-10 & 1.497E-09 & 2.631E-09 & 4.597E-09 &           \\
4  & 5.063E-12 & 2.013E-11 & 5.374E-11 & 1.203E-10 & 2.443E-10 & 4.664E-10 & 8.558E-10 & 1.530E-09 & 2.694E-09 & 4.716E-09 &           \\
5  & 5.207E-12 & 2.073E-11 & 5.539E-11 & 1.242E-10 & 2.523E-10 & 4.823E-10 & 8.858E-10 & 1.586E-09 & 2.797E-09 &           &           \\
6  & 5.412E-12 & 2.157E-11 & 5.768E-11 & 1.294E-10 & 2.632E-10 & 5.037E-10 & 9.263E-10 & 1.661E-09 & 2.934E-09 &           &           \\
\hline\hline
\end{tabular}
\end{center}
\end{sidewaystable}

\begin{sidewaystable}
\begin{center}
\caption{Averaged decay rates $\bar{A}_{v,L,v',L'}/(B_0^4 r_c^2)$ in s$^{-1}$T$^{-4}$mm$^{-2}$ with $v'=v-1$, $L'=L+2$. Stars indicate cases where the $(v-1,L+2)$ level lies above
the $(v,L)$ level. In these cases, the ''reversed'' decay rate $\bar{A}_{v',L',v,L}/(B_0^4 r_c^2)$ is given instead.
\label{A-dlp2-dv1}}
\begin{tabular}{c@{\hspace{2mm}}l@{\hspace{2mm}}l@{\hspace{2mm}}l@{\hspace{2mm}}l@{\hspace{2mm}}l@{\hspace{2mm}}l@{\hspace{2mm}}l@{\hspace{2mm}}l@{\hspace{2mm}}l@{\hspace{2mm}}l@{\hspace{2mm}}l@{\hspace{1mm}}l}
\hline\hline
$L/v$ & \makecell[c]{1} & \makecell[c]{2} & \makecell[c]{3} & \makecell[c]{4} & \makecell[c]{5} & \makecell[c]{6} & \makecell[c]{7} & \makecell[c]{8} & \makecell[c]{9} & \makecell[c]{10} & \makecell[c]{11} & \makecell[c]{12} \\
\hline
 0 & 1.056E-10 & 2.798E-10 & 5.574E-10 & 9.913E-10 & 1.663E-09 & 2.701E-09 & 4.312E-09 & 6.835E-09 & 1.085E-08 & 1.739E-08 & 2.835E-08 & 4.750E-08 \\
 1 & 5.695E-11 & 1.511E-10 & 3.013E-10 & 5.362E-10 & 9.001E-10 & 1.462E-09 & 2.335E-09 & 3.701E-09 & 5.875E-09 & 9.412E-09 & 1.534E-08 & 2.570E-08 \\
 2 & 4.364E-11 & 1.159E-10 & 2.312E-10 & 4.117E-10 & 6.913E-10 & 1.123E-09 & 1.794E-09 & 2.843E-09 & 4.513E-09 & 7.230E-09 & 1.178E-08 & 1.974E-08 \\
 3 & 3.590E-11 & 9.539E-11 & 1.904E-10 & 3.391E-10 & 5.695E-10 & 9.255E-10 & 1.478E-09 & 2.342E-09 & 3.717E-09 & 5.953E-09 & 9.702E-09 &           \\
 4 & 3.024E-11 & 8.036E-11 & 1.604E-10 & 2.857E-10 & 4.798E-10 & 7.796E-10 & 1.244E-09 & 1.972E-09 & 3.129E-09 & 5.009E-09 & 8.162E-09 &           \\
 5 & 2.566E-11 & 6.820E-11 & 1.361E-10 & 2.424E-10 & 4.069E-10 & 6.609E-10 & 1.054E-09 & 1.670E-09 & 2.648E-09 & 4.237E-09 &           &           \\
 6 & 2.179E-11 & 5.788E-11 & 1.155E-10 & 2.055E-10 & 3.448E-10 & 5.596E-10 & 8.921E-10 & 1.412E-09 & 2.236E-09 & 3.573E-09 &           &           \\
\hline\hline
\end{tabular}
\end{center}
\end{sidewaystable}

\begin{sidewaystable}
\begin{center}
\caption{Same as Table~\ref{A-dlp2-dv1}, with $v'=v-2$.
\label{A-dlp2-dv2}}
\begin{tabular}{c@{\hspace{2mm}}l@{\hspace{2mm}}l@{\hspace{2mm}}l@{\hspace{2mm}}l@{\hspace{2mm}}l@{\hspace{2mm}}l@{\hspace{2mm}}l@{\hspace{2mm}}l@{\hspace{2mm}}l@{\hspace{2mm}}l@{\hspace{1mm}}l}
\hline\hline
$L/v$ & \makecell[c]{2} & \makecell[c]{3} & \makecell[c]{4} & \makecell[c]{5} & \makecell[c]{6} & \makecell[c]{7} & \makecell[c]{8} & \makecell[c]{9} & \makecell[c]{10} & \makecell[c]{11} & \makecell[c]{12} \\
\hline
 0 & 1.392E-11 & 5.119E-11 & 1.269E-10 & 2.654E-10 & 5.057E-10 & 9.117E-10 & 1.588E-09 & 2.712E-09 & 4.587E-09 & 7.753E-09 & 1.322E-08 \\
 1 & 9.373E-12 & 3.450E-11 & 8.562E-11 & 1.792E-10 & 3.417E-10 & 6.165E-10 & 1.075E-09 & 1.838E-09 & 3.114E-09 & 5.275E-09 & 9.023E-09 \\
 2 & 8.947E-12 & 3.298E-11 & 8.194E-11 & 1.717E-10 & 3.277E-10 & 5.918E-10 & 1.033E-09 & 1.769E-09 & 3.003E-09 & 5.102E-09 & 8.759E-09 \\
 3 & 9.163E-12 & 3.383E-11 & 8.416E-11 & 1.765E-10 & 3.374E-10 & 6.102E-10 & 1.067E-09 & 1.830E-09 & 3.115E-09 & 5.307E-09 &           \\
 4 & 9.614E-12 & 3.555E-11 & 8.858E-11 & 1.861E-10 & 3.561E-10 & 6.451E-10 & 1.130E-09 & 1.943E-09 & 3.315E-09 & 5.668E-09 &           \\
 5 & 1.018E-11 & 3.772E-11 & 9.415E-11 & 1.981E-10 & 3.798E-10 & 6.891E-10 & 1.210E-09 & 2.085E-09 & 3.568E-09 &           &           \\
 6 & 1.083E-11 & 4.018E-11 & 1.004E-10 & 2.117E-10 & 4.066E-10 & 7.393E-10 & 1.301E-09 & 2.249E-09 & 3.862E-09 &           &           \\
\hline\hline
\end{tabular}
\end{center}
\end{sidewaystable}

\begin{sidewaystable}
\begin{center}
\caption{Total decay rates $\bar{A}_{v,L}/(B_0^4 r_c^2)$ in s$^{-1}$T$^{-4}$mm$^{-2}$.
\label{A-total}}
\begin{tabular}{c@{\hspace{2mm}}l@{\hspace{2mm}}l@{\hspace{2mm}}l@{\hspace{2mm}}l@{\hspace{2mm}}l@{\hspace{2mm}}l@{\hspace{2mm}}l@{\hspace{2mm}}l@{\hspace{2mm}}l@{\hspace{2mm}}l@{\hspace{2mm}}l@{\hspace{2mm}}l@{\hspace{1mm}}l}
\hline\hline
$L/v$ & \makecell[c]{0} & \makecell[c]{1} & \makecell[c]{2} & \makecell[c]{3} & \makecell[c]{4} & \makecell[c]{5} & \makecell[c]{6} & \makecell[c]{7} & \makecell[c]{8} & \makecell[c]{9} & \makecell[c]{10} & \makecell[c]{11} & \makecell[c]{12} \\
\hline
0  & 0         & 2.353E-10 & 5.987E-10 & 1.157E-09 & 2.013E-09 & 3.327E-09 & 5.354E-09 & 8.503E-09 & 1.346E-08 & 2.136E-08 & 3.427E-08 & 5.592E-08 & 9.367E-08 \\
1  & 0         & 2.363E-10 & 6.013E-10 & 1.162E-09 & 2.022E-09 & 3.343E-09 & 5.381E-09 & 8.549E-09 & 1.353E-08 & 2.150E-08 & 3.450E-08 & 5.633E-08 & 9.442E-08 \\
2  & 6.693E-14 & 2.384E-10 & 6.067E-10 & 1.173E-09 & 2.042E-09 & 3.377E-09 & 5.438E-09 & 8.644E-09 & 1.369E-08 & 2.177E-08 & 3.497E-08 & 5.717E-08 & 9.599E-08 \\
3  & 3.996E-13 & 2.420E-10 & 6.156E-10 & 1.190E-09 & 2.073E-09 & 3.430E-09 & 5.528E-09 & 8.795E-09 & 1.395E-08 & 2.220E-08 & 3.571E-08 & 5.850E-08 &           \\
4  & 1.224E-12 & 2.475E-10 & 6.286E-10 & 1.216E-09 & 2.118E-09 & 3.507E-09 & 5.656E-09 & 9.010E-09 & 1.431E-08 & 2.281E-08 & 3.678E-08 & 6.041E-08 &           \\
5  & 2.774E-12 & 2.553E-10 & 6.466E-10 & 1.250E-09 & 2.178E-09 & 3.610E-09 & 5.830E-09 & 9.299E-09 & 1.479E-08 & 2.364E-08 & 3.822E-08 &           &           \\
6  & 5.299E-12 & 2.659E-10 & 6.702E-10 & 1.295E-09 & 2.257E-09 & 3.744E-09 & 6.054E-09 & 9.674E-09 & 1.542E-08 & 2.471E-08 & 4.010E-08 &           &           \\
\hline\hline
\end{tabular}
\end{center}
\end{sidewaystable}

\end{document}


\title{Supplemental Material: Quenching of excited rovibrational states of H$_2^+$ in an external magnetic field}

\author{Jean-Philippe Karr}
\affiliation{Laboratoire Kastler Brossel, Sorbonne Universit\'e, CNRS, ENS-PSL Research University, Coll\`ege de France\\
4 place Jussieu, F-75005 Paris, France}
\affiliation{Universit\'e d'Evry-Val d'Essonne, Universit\'e Paris-Saclay, Boulevard Fran\c cois Mitterrand, F-91000 Evry, France}

\maketitle

The two-photon matrix elements $Q_s$, $Q_v$, and $Q_t$ defined in Eqs. (21-23), in atomic units (that is $(ea_0)^2/E_h = 4\pi\epsilon_0 a_0^3$) where $a_0$ is the Bohr radius and $E_h$ the Hartree energy), are given in the following Tables for $L \to L' = L-2$ ($\Delta v = v - v' = 0,1,2$, Tables ~\ref{q-dlm2-dv0}-\ref{q-dlm2-dv2}) , $L \to L' = L$ ($\Delta v = 1,2$, Tables~\ref{q-dl0-dv1}-\ref{q-dl0-dv2}), and $L \to L' = L+2$ transitions ($\Delta v = 1,2$, Tables~\ref{q-dlp2-dv1}-\ref{q-dlp2-dv2}).

It is worth noting that for $L \to L$ transitions, the rank 1 component $Q_v$ is comparatively small and may safely be ignored for calculating decay rates at the 4-digit accuracy level.

\begin{table}
\begin{center}
\caption{Two-photon matrix elements $Q_t$ in atomic units, with $L' = L-2$ and $v' = v$.
\label{q-dlm2-dv0}}
\begin{tabular}{c@{\hspace{2mm}}l@{\hspace{2mm}}l@{\hspace{2mm}}l@{\hspace{2mm}}l@{\hspace{2mm}}l@{\hspace{2mm}}l@{\hspace{2mm}}l@{\hspace{2mm}}l@{\hspace{2mm}}l@{\hspace{2mm}}l@{\hspace{2mm}}l@{\hspace{2mm}}l@{\hspace{1mm}}l}
\hline\hline
$L/v$ & \makecell[c]{0} & \makecell[c]{1} & \makecell[c]{2} & \makecell[c]{3} & \makecell[c]{4} & \makecell[c]{5} & \makecell[c]{6} & \makecell[c]{7} & \makecell[c]{8} & \makecell[c]{9} & \makecell[c]{10} & \makecell[c]{11} & \makecell[c]{12} \\
\hline
 2 & 2.6840 & 3.8218 & 5.3437 & 7.3899 & 10.163 & 13.962 & 19.240 & 26.705 & 37.504 & 53.572 & 78.347 & 118.33 & 186.85 \\
 3 & 2.0986 & 2.9873 & 4.1762 & 5.7752 & 7.9431 & 10.915 & 15.047 & 20.896 & 29.368 & 41.992 & 61.494 & 93.049 &        \\
 4 & 1.9704 & 2.8034 & 3.9182 & 5.4183 & 7.4535 & 10.246 & 14.132 & 19.643 & 27.637 & 39.578 & 58.077 & 88.125 &        \\
 5 & 1.9316 & 2.7466 & 3.8376 & 5.3068 & 7.3018 & 10.042 & 13.863 & 19.290 & 27.182 & 39.006 & 57.398 &        &        \\
 6 & 1.9313 & 2.7440 & 3.8328 & 5.3003 & 7.2954 & 10.040 & 13.874 & 19.333 & 27.297 & 39.275 & 58.000 &        &        \\
 7 & 1.9542 & 2.7744 & 3.8740 & 5.3575 & 7.3776 & 10.162 & 14.060 & 19.628 & 27.780 & 40.101 &        &        &        \\
 8 & 1.9946 & 2.8292 & 3.9493 & 5.4626 & 7.5267 & 10.378 & 14.382 & 20.120 & 28.562 & 41.393 &        &        &        \\
\hline\hline
\end{tabular}
\end{center}
\end{table}
\clearpage

\begin{table}
\begin{center}
\caption{Same as Table~\ref{q-dlm2-dv0}, with $v' = v-1$.
\label{q-dlm2-dv1}}
\begin{tabular}{c@{\hspace{2mm}}l@{\hspace{2mm}}l@{\hspace{2mm}}l@{\hspace{2mm}}l@{\hspace{2mm}}l@{\hspace{2mm}}l@{\hspace{2mm}}l@{\hspace{2mm}}l@{\hspace{2mm}}l@{\hspace{2mm}}l@{\hspace{2mm}}l@{\hspace{1mm}}l}
\hline\hline
$L/v$ & \makecell[c]{1} & \makecell[c]{2} & \makecell[c]{3} & \makecell[c]{4} & \makecell[c]{5} & \makecell[c]{6} & \makecell[c]{7} & \makecell[c]{8} & \makecell[c]{9} & \makecell[c]{10} & \makecell[c]{11} & \makecell[c]{12} \\
\hline
 2 & 1.1077  & 1.9670 & 3.0372 & 4.4466 & 6.3497 & 8.9678 & 12.636 & 17.884 & 25.581 & 37.225 & 55.541 & 85.847 \\
 3 & 0.83969 & 1.4893 & 2.2973 & 3.3606 & 4.7959 & 6.7703 & 9.5370 & 13.496 & 19.307 & 28.103 & 41.956 &        \\
 4 & 0.76393 & 1.3533 & 2.0856 & 3.0487 & 4.3487 & 6.1374 & 8.6453 & 12.237 & 17.514 & 25.517 & 38.147 &        \\
 5 & 0.72540 & 1.2836 & 1.9764 & 2.8874 & 4.1172 & 5.8104 & 8.1865 & 11.594 & 16.610 & 24.235 &        &        \\
 6 & 0.70238 & 1.2415 & 1.9101 & 2.7892 & 3.9765 & 5.6127 & 7.9121 & 11.216 & 16.092 & 23.527 &        &        \\
 7 & 0.68829 & 1.2153 & 1.8686 & 2.7276 & 3.8889 & 5.4911 & 7.7471 & 10.997 & 15.807 &        &        &        \\
 8 & 0.68043 & 1.2003 & 1.8445 & 2.6920 & 3.8389 & 5.4241 & 7.6612 & 10.893 & 15.697 &        &        &        \\
\hline\hline
\end{tabular}
\end{center}
\end{table}
\clearpage

\begin{table}
\begin{center}
\caption{Same as Table~\ref{q-dlm2-dv0}, with $v' = v-2$.
\label{q-dlm2-dv2}}
\begin{tabular}{c@{\hspace{2mm}}l@{\hspace{2mm}}l@{\hspace{2mm}}l@{\hspace{2mm}}l@{\hspace{2mm}}l@{\hspace{2mm}}l@{\hspace{2mm}}l@{\hspace{2mm}}l@{\hspace{2mm}}l@{\hspace{2mm}}l@{\hspace{1mm}}l}
\hline\hline
$L/v$ & \makecell[c]{2} & \makecell[c]{3} & \makecell[c]{4} & \makecell[c]{5} & \makecell[c]{6} & \makecell[c]{7} & \makecell[c]{8} & \makecell[c]{9} & \makecell[c]{10} & \makecell[c]{11} & \makecell[c]{12} \\
\hline
 2 & 0.11149 & 0.23388 & 0.40416 & 0.64377 & 0.98330 & 1.4686  & 2.1704  & 3.2013 & 4.7471 & 7.1302 & 10.946 \\
 3 & 0.07752 & 0.16263 & 0.28110 & 0.44791 & 0.68441 & 1.0226  & 1.5116  & 2.2296 & 3.3057 & 4.9630 &        \\
 4 & 0.06428 & 0.13492 & 0.23337 & 0.37214 & 0.56908 & 0.85087 & 1.2586  & 1.8572 & 2.7541 & 4.1349 &        \\
 5 & 0.05525 & 0.11610 & 0.20104 & 0.32098 & 0.49144 & 0.73564 & 1.0892  & 1.6086 & 2.3868 &        &        \\
 6 & 0.04807 & 0.10116 & 0.17547 & 0.28066 & 0.43045 & 0.64536 & 0.95687 & 1.4147 & 2.1008 &        &        \\
 7 & 0.04196 & 0.08852 & 0.15392 & 0.24676 & 0.37933 & 0.56989 & 0.84645 & 1.2532 &        &        &        \\
 8 & 0.03660 & 0.07745 & 0.13512 & 0.21730 & 0.33499 & 0.50456 & 0.75098 & 1.1136 &        &        &        \\
\hline\hline
\end{tabular}
\end{center}
\end{table}
\clearpage

\begin{center}
{\scriptsize
\begin{longtable}{c@{\hspace{1.5mm}}r@{\hspace{1.5mm}}r@{\hspace{1.5mm}}r@{\hspace{1.5mm}}r@{\hspace{1.5mm}}r@{\hspace{1.5mm}}r@{\hspace{1.5mm}}r@{\hspace{1.5mm}}r@{\hspace{1.5mm}}r@{\hspace{1.5mm}}r@{\hspace{1.5mm}}r@{\hspace{1.5mm}}r}
\caption{Two-photon matrix elements $Q_s$ (first line) $Q_v$ (second line) and $Q_t$ (third line) in atomic units, with $L'=L$ and $v'=v-1$. \label{q-dl0-dv1}} \\
\hline\hline
$L/v$ & 1    & 2        & 3        & 4        & 5        & 6        & 7        & 8        & 9        & 10       & 11       & 12       \\
\hline
\endfirsthead
\multicolumn{13}{c}
{\tablename\ \thetable\ (continued)} \\
\hline
$L/v$ & 1    & 2        & 3        & 4        & 5        & 6        & 7        & 8        & 9        & 10       & 11       & 12       \\
\hline
\endhead
\hline \multicolumn{13}{c}{(continued on next page)} \\
\endfoot
\hline\hline
\endlastfoot

0  & -0.83797 & -1.4013  & -2.0514  & -2.8653  & -3.9272  & -5.3537  & -7.3199  & -10.102  & -14.153  & -20.255  & -29.827  & -45.641  \\
1  & -0.84102 & -1.4067  & -2.0598  & -2.8778  & -3.9456  & -5.3805  & -7.3596  & -10.162  & -14.245  & -20.401  & -30.070  & -46.067  \\
1  &  0.00004 &  0.00009 &  0.00016 &  0.00028 &  0.00047 &  0.00079 &  0.00134 &  0.00233 &  0.00415 &  0.00772 &  0.01520 &  0.03240 \\
1  &  0.73131 &  1.3010  &  2.0122  &  2.9502  &  4.2184  &  5.9648  &  8.4139  &  11.920  &  17.068  &  24.861  &  37.131  &  57.453  \\
2  & -0.84714 & -1.4175  & -2.0766  & -2.9029  & -3.9824  & -5.4346  & -7.4397  & -10.282  & -14.431  & -20.697  & -30.562  & -46.933  \\
2  &  0.00008 &  0.00016 &  0.00029 &  0.00049 &  0.00083 &  0.00140 &  0.00237 &  0.00411 &  0.00734 &  0.01369 &  0.02708 &  0.05797 \\
2  &  0.62401 &  1.1104  &  1.7178  &  2.5194  &  3.6038  &  5.0984  &  7.1964  &  10.204  &  14.624  &  21.329  &  31.910  &  49.487  \\
3  & -0.85639 & -1.4339  & -2.1020  & -2.9409  & -4.0383  & -5.5167  & -7.5615  & -10.466  & -14.715  & -21.150  & -31.317  &          \\
\end{longtable}}
\end{center}
\clearpage

\begin{center}
{\scriptsize
\begin{longtable}{c@{\hspace{1.5mm}}r@{\hspace{1.5mm}}r@{\hspace{1.5mm}}r@{\hspace{1.5mm}}r@{\hspace{1.5mm}}r@{\hspace{1.5mm}}r@{\hspace{1.5mm}}r@{\hspace{1.5mm}}r@{\hspace{1.5mm}}r@{\hspace{1.5mm}}r@{\hspace{1.5mm}}r@{\hspace{1.5mm}}r}
\caption{Same as Table~\ref{q-dl0-dv1}, with $v'=v-2$. \label{q-dl0-dv2}} \\
\hline\hline
$L/v$ & 2 & 3 & 4 & 5 & 6 & 7 & 8 & 9 & 10 & 11 & 12 \\
\hline
\endfirsthead
\multicolumn{12}{c}
{\tablename\ \thetable\ (continued)} \\
\hline
$L/v$ & 2 & 3 & 4 & 5 & 6 & 7 & 8 & 9 & 10 & 11 & 12 \\
\hline
\endhead
\hline \multicolumn{12}{c}{(continued on next page)} \\
\endfoot
\hline\hline
\endlastfoot

 0 & -0.03562 & -0.08270 & -0.15560 & -0.26641 & -0.43284 & -0.68172 & -1.0547  & -1.6186  & -2.4838  & -3.8426  & -6.0497  \\
 1 & -0.03587 & -0.08328 & -0.15668 & -0.26825 & -0.43585 & -0.68654 & -1.0624  & -1.6308  & -2.5035  & -3.8751  & -6.1053  \\
 1 &  0.00003 &  0.00006 &  0.00012 &  0.00023 &  0.00041 &  0.00073 &  0.00131 &  0.00240 &  0.00453 &  0.00895 &  0.01885 \\
 1 &  0.08291 &  0.17403 &  0.30084 &  0.47932 &  0.73230 &  1.0941  &  1.6179  &  2.3884  &  3.5465  &  5.3373  &  8.2150  \\
 2 & -0.03639 & -0.08444 & -0.15884 & -0.27195 & -0.44191 & -0.69626 & -1.0779  & -1.6554  & -2.5433  & -3.9410  & -6.2183  \\
 2 &  0.00005 &  0.00011 &  0.00022 &  0.00040 &  0.00072 &  0.00129 &  0.00232 &  0.00425 &  0.00804 &  0.01592 &  0.03367 \\
 2 &  0.07078 &  0.14860 &  0.25698 &  0.40961 &  0.62610 &  0.93596 &  1.3849  &  2.0461  &  3.0412  &  4.5827  &  7.0652  \\
 3 & -0.03717 & -0.08621 & -0.16213 & -0.27758 & -0.45113 & -0.71106 & -1.1014  & -1.6931  & -2.6042  & -4.0419  &          \\
\end{longtable}}
\end{center}

\begin{table}
\begin{center}
\caption{Two-photon matrix elements $Q_t$ in atomic units, with $L'=L+2$ and $v'=v-1$. Stars indicate cases where the $(v-1,L+2)$ level lies above
the $(v,L)$ level. In these cases, the ''reversed'' matrix element $\frac{\left\langle v L || Q^{(2)} || v-1 L+2 \right\rangle}{\sqrt{2L+5}}
=\sqrt{\frac{2L+5}{2L+1}} \frac{\left\langle v-1 L+2 || Q^{(2)} || v L \right\rangle}{\sqrt{2L+1}}$ is given instead.
\label{q-dlp2-dv1}}
\begin{tabular}{c@{\hspace{2mm}}l@{\hspace{2mm}}l@{\hspace{2mm}}l@{\hspace{2mm}}l@{\hspace{2mm}}l@{\hspace{2mm}}l@{\hspace{2mm}}l@{\hspace{2mm}}l@{\hspace{2mm}}l@{\hspace{2mm}}l@{\hspace{2mm}}l@{\hspace{1mm}}l}
\hline\hline
$L/v$ & \makecell[c]{1} & \makecell[c]{2} & \makecell[c]{3} & \makecell[c]{4} & \makecell[c]{5} & \makecell[c]{6} & \makecell[c]{7} & \makecell[c]{8} & \makecell[c]{9} & \makecell[c]{10} & \makecell[c]{11} & \makecell[c]{12} \\
\hline
 0 & 0.54126 & 0.96458 & 1.4942 & 2.1938 & 3.1409 & 4.4468 & 6.2804 & 8.9088 & 12.772 & 18.631 & 27.872 & 43.213 \\
 1 & 0.63690 & 1.1363  & 1.7622 & 2.5902 & 3.7126 & 5.2622 & 7.4413 & 10.570 & 15.179 & 22.185 & 33.270 & 51.744 \\
 2 & 0.69918 & 1.2489  & 1.9390 & 2.8533 & 4.0948 & 5.8118 & 8.2309 & 11.713 & 16.854 & 24.697 & 37.156 & 58.034 \\
 3 & 0.75249 & 1.3456  & 2.0916 & 3.0818 & 4.4287 & 6.2955 & 8.9318 & 12.737 & 18.375 & 27.011 & 40.806 &        \\
 4 & 0.80403 & 1.4394  & 2.2401 & 3.3050 & 4.7567 & 6.7735 & 9.6299 & 13.766 & 19.922 & 29.399 & 44.638 &        \\
 5 & 0.85695 & 1.5359  & 2.3933 & 3.5361 & 5.0978 & 7.2736 & 10.365 & 14.861 & 21.584 & 32.000 &        &        \\
 6 & 0.91303 & 1.6383  & 2.5562 & 3.7827 & 5.4636 & 7.8129 & 11.164 & 16.060 & 23.425 & 34.920 &        &        \\
\hline\hline
\end{tabular}
\end{center}
\end{table}

\begin{table}
\begin{center}
\caption{Same as Table~\ref{q-dlp2-dv1}, with $v'=v-2$.
\label{q-dlp2-dv2}}
\begin{tabular}{c@{\hspace{2mm}}l@{\hspace{2mm}}l@{\hspace{2mm}}l@{\hspace{2mm}}l@{\hspace{2mm}}l@{\hspace{2mm}}l@{\hspace{2mm}}l@{\hspace{2mm}}l@{\hspace{2mm}}l@{\hspace{2mm}}l@{\hspace{1mm}}l}
\hline\hline
$L/v$ & \makecell[c]{2} & \makecell[c]{3} & \makecell[c]{4} & \makecell[c]{5} & \makecell[c]{6} & \makecell[c]{7} & \makecell[c]{8} & \makecell[c]{9} & \makecell[c]{10} & \makecell[c]{11} & \makecell[c]{12} \\
\hline
 0 & 0.06830 & 0.14349 & 0.24824 & 0.39579 & 0.60510 & 0.90477 & 1.3392 & 1.9796 & 2.9447 & 4.4422 & 6.8591 \\
 1 & 0.08581 & 0.18045 & 0.31245 & 0.49859 & 0.76297 & 1.1420  & 1.6924 & 2.5056 & 3.7347 & 5.6488 & 8.7521 \\
 2 & 0.10014 & 0.21081 & 0.36541 & 0.58374 & 0.89434 & 1.3405  & 1.9898 & 2.9517 & 4.4107 & 6.6931 & 10.415 \\
 3 & 0.11409 & 0.24048 & 0.41737 & 0.66764 & 1.0244  & 1.5379  & 2.2874 & 3.4016 & 5.0986 & 7.7679 &        \\
 4 & 0.12855 & 0.27133 & 0.47161 & 0.75556 & 1.1613  & 1.7469  & 2.6045 & 3.8845 & 5.8440 & 8.9460 &        \\
 5 & 0.14396 & 0.30433 & 0.52981 & 0.85030 & 1.3095  & 1.9745  & 2.9520 & 4.4180 & 6.6755 &        &        \\
 6 & 0.16062 & 0.34011 & 0.59317 & 0.95390 & 1.4724  & 2.2261  & 3.3389 & 5.0170 & 7.6191 &        &        \\
\hline\hline
\end{tabular}
\end{center}
\end{table}